\documentclass[twocolumn]{jpsj3}
\usepackage{txfonts}

\title{
Multiplicative Modeling of Children's Growth and Its Statistical Properties
}

\author{Hiroto Kuninaka$^1$\thanks{E-mail: kuninaka@edu.mie-u.ac.jp} 
and \name{Mitsugu Matsushita}$^2$}
\inst{$^1$Faculty of Education, Mie University, 1577 Kurima-machiya-cho, 
Tsu, Mie 514-8507, Japan \\
$^2$ Meiji Institute for Advanced Study of Mathematical Sciences, 
4-21-1 Nakano, Nakano-ku, Tokyo 164-8525, Japan}

\abst{
We develop a numerical growth model that can predict the statistical properties 
of the height distribution of Japanese children. Our previous studies 
have clarified that the height distribution of schoolchildren shows a 
transition from 
the lognormal distribution to the normal distribution during puberty. 
In this study, we demonstrate by simulation 
that the transition occurs owing to the variability 
of the onset of puberty. 
}

\begin{document}
\maketitle

\section{Introduction}

\begin{figure}[h]
\begin{center}
\includegraphics[width=0.4\textwidth]{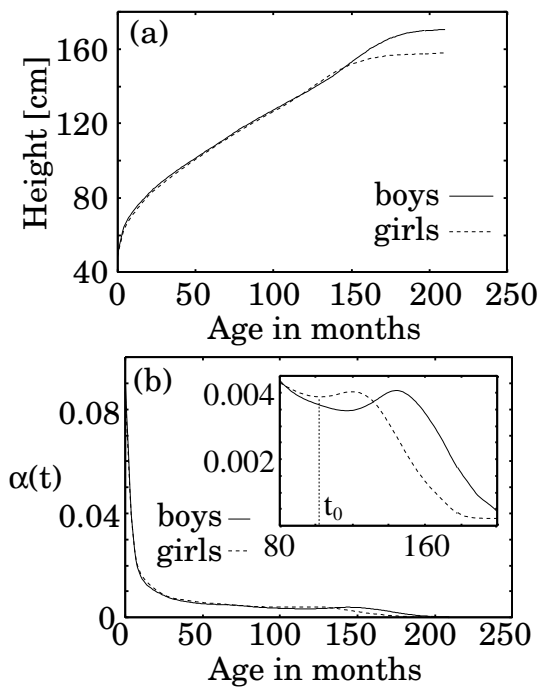}
\end{center}
\caption{
(a) Growth curves for male (solid) and female (broken) children. 
(b) Growth rate defined using Eq. (\ref{gr}). The growth rate of 
female children has a local minimum at $t_{0}=102$ months. 
}
\label{fig1}
\end{figure}

The growth process of children is often characterized 
by the time evolution of their height in addition to 
those of their weight and sitting height. 
Although genetic factors affect the growth process and final adult 
height the most, the socioeconomic position of a family, nutrition, 
and diseases are also important factors\cite{silventoinen}. 
As an example of children's growth, we show the growth curves of male (solid) 
and female (broken) Japanese children\cite{syoni} in Fig. \ref{fig1}(a).  
The shape of the curves shown in Fig. \ref{fig1}(a) is typical of human growth 
and can also be observed in the case of U.S. children\cite{uschildren}. 

In medicine, various mathematical models describing the average growth 
of children have been proposed\cite{preece,jolicoeur,kanefuji,canessa}, 
mainly for the purpose of medication in children with unusual growth. 
For example, Preece and Baines developed a five-parameter growth 
model as follows:
\begin{equation}\label{PB}
H(t) = U - \frac{2(U-E)}{\exp[A(t-C)] + \exp[B(t-C)]},
\end{equation}
where $H(t)$ is the average height (cm) of children at age $t$ 
(month). The parameters $(A, B, C, E, U)$ are called the growth 
parameters of the model, which have biological significance, such as 
the final adult height ($U$). 
Kanefuji has shown that the growth curve of Japanese boys born in 1962 
can be well approximated by Eq. (\ref{PB}) with appropriate growth parameters
\cite{kanefuji_tokei}.

In general, the growth rate of a newborn baby shows a monotonic decrease 
until the onset of the growth spurt, after which the growth rate increases to 
a local maximum followed by a monotonic decrease to zero. Here, we define  
the average growth rate $\alpha(t)$ by  
\begin{equation}\label{gr}
\alpha(t) \equiv \frac{H(t + \Delta t)-H(t)}{H(t)}, 
\end{equation}
where $\Delta t = 1$ (month). 
Figure \ref{fig1}(b) shows the average growth rate calculated using Eq.(\ref{gr}) 
from the data of the Japanese Society for Pediatric Endocrinology\cite{syoni},  
where the solid and broken curves represent the $\alpha(t)$ values of data 
for boys and girls, respectively. 
The data are smoothed by the Bezier interpolation.  
As can be seen in the inset of Fig. \ref{fig1}(b), 
each curve has a local minimum around $100-120$ months old, which can be 
defined as the onset of a growth spurt. For example, the growth spurt begins 
at $t_{0}=102$ months in the case of girls.

Here, we should distinguish between puberty and growth spurts. 
Puberty is the period during which children's bodies become  
adult bodies capable of reproduction. In the first half of puberty, 
the growth rate of children positively accelerates 
towards a local maximum value followed by a monotonic decrease to zero. 
The period of the positive acceleration of growth is called a growth spurt. 
Japanese girls begin puberty at ages 9-10,  which continues up to ages 15-17, 
during which they experience menarche around the age of 12 on average. 
On the other hand, boys begin puberty at ages 11-12, which continues up to 
ages 16-17. 

In addition to the average growth, the height distribution is often useful 
for the evaluation of children's development in a given group
\cite{tokushima, ahearn, schilling}. 
In general, the height and weight distributions are believed to obey 
the normal distribution. However, some studies have shown that they approximately 
obey the lognormal distribution\cite{soltow,limpert}. 
In our previous works\cite{kuninaka2009, mitsuhashi, kagaku, JPSJinvited}, 
we investigated the height distribution of Japanese children, 
the ages of which range from 5 to 17,  
on the basis of  the data from the Ministry of Education, Culture, Sports, Science 
and Technology\cite{mext}. 
Our findings are summarized as follows: 
(i) the height distribution of Japanese schoolchildren obeys the lognormal 
distribution before puberty, (ii) the height distribution shows a transition 
to the normal distribution during puberty, and (iii) the height distribution 
fits the lognormal distribution equally well as the normal distribution after puberty.  
However, the mechanism of the transition is still unclear. 


The aim of this study is to clarify the origin of the transition 
of the height distribution during puberty  with a growth model constructed 
on a biological basis. 
The organization of this paper is as follows. 
In the next section, we will introduce our growth model. 
We will show the simulation results by our growth model in Sect. 3. 
We will devote Sect. 4 to the discussion of our results. 
In Sect. 5, we will summarize our results.  

\section{Model}

Let us introduce our growth model. Some studies have implied that 
the growth process for living organisms is multiplicative 
from the fact that the body size distribution often obeys the 
lognormal distribution
\cite{huxley, canessa, silventoinen}. 
From Eq. (\ref{gr}), we describe the children's growth 
by the multiplicative process as 
\begin{equation}\label{mp}
H^{(i)}(t + \Delta t) = (1 + \alpha^{(i)}(t)) H^{(i)}(t).
\end{equation}
Here, $H^{(i)}(t)$ and $\alpha^{(i)}(t)$ are the height and 
growth rate of the $i$-th body at age $t$ (months), respectively. 
We use $1$ month for $\Delta t$. The total number of growing bodies studied 
is $10^{6}$. The initial height of the $i$-th body, $H^{(i)}(0)$, is randomly 
chosen from the lognormal distribution of the height $x$, 
\begin{equation}\label{ini}
f(x) = \frac{1}{\sqrt{2 \pi} \sigma x} \exp
\left[
- \frac{(\log x - \mu)^{2}}{2 \sigma^{2}}
\right], 
\end{equation}
with $\mu=3.878$ and $\sigma = 2.15$\cite{suku}.

\begin{figure}[h]
\begin{center}
\includegraphics[width=0.4\textwidth]{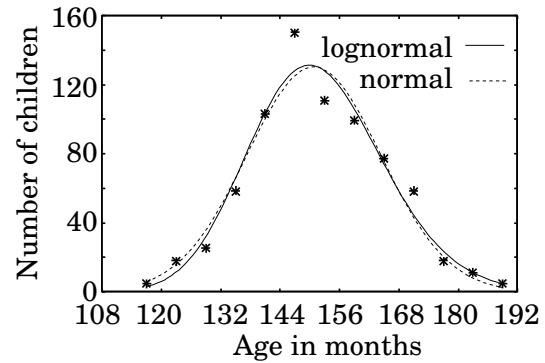}
\end{center}
\caption{
Distribution of onset age (in month) of menarche. 
Solid and broken curves are curves fitted 
by lognormal and normal distributions, respectively. 
}
\label{fig2}
\end{figure}

We define $\alpha^{(i)}(t)$ on the basis of the average growth rate 
$\alpha(t)$ of girls, which is represented by the broken curve 
in Fig. \ref{fig1}(b). 
The reason why we use the average growth rate of girls is that 
the puberty of girls is clearly characterized by menarche, 
which is statistically examined. 
We introduce two kinds of fluctuation in $\alpha(t)$ as follows. 
First, we give the variability in the onset of growth spurt. 
The puberty of girls is characterized by the onset of menarche,  
which occurs around  ages 144-156 months. 
Figure \ref{fig2} shows the distribution of the onset of menarche 
in Japan\cite{menarche},  
where the data are well approximated by both the normal and lognormal distributions. 
The broken curve shows the best-fit normal distribution with a mean of $145.8$ months 
and a standard deviation of $13.56$ months,  
while the solid curve shows the best-fit lognormal distribution. 
Assuming that the onset of the growth spurt has a close relationship with 
that of menarche and obeys the normal distribution, 
we choose a normal random number from the normal 
distribution with a mean of $132$ months and a standard deviation of 
$12.8$ months to define the onset of the growth spurt $t_{s}^{(i)}$ for each body. 

After $t_{s}^{(i)}$ is chosen, we define the function $\alpha^{\ast (i)}(t)$ such that 
the following relation is fulfilled:
\begin{equation}
\alpha^{\ast (i)}\left(\tilde{t}\right) = \alpha(t),
\end{equation}
where $\tilde{t}$ is the scaled  age, $\tilde{t} \equiv t \times (t_{s}^{(i)}/t_{0})$.
Next, we give a fluctuation in $\alpha^{\ast (i)}\left( \tilde{t} \right)$ as  
\begin{equation}
\alpha^{(i)}\left( \tilde{t} \right) = \alpha^{\ast (i)}\left( \tilde{t} \right) + \delta \alpha^{(i)}\left( \tilde{t} \right),
\end{equation}
where $\delta \alpha^{(i)}\left( \tilde{t} \right)$ is randomly chosen from the normal distribution 
with a mean of $0$ and a standard deviation of $\sigma = 10^{-3}$.



\section{Results}

\begin{figure}[h]
\begin{center}
\includegraphics[width=0.4\textwidth]{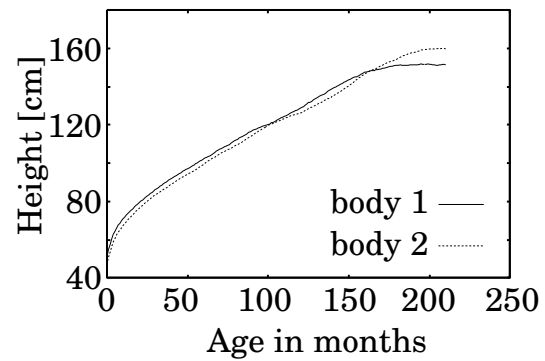}
\end{center}
\caption{
Time evolution of individual heights.
}
\label{fig4}
\end{figure}

First we show the time evolution of the heights of two arbitrarily chosen bodies 
(called bodies 1 and 2) in Fig. \ref{fig4}, each of which shows a similar growth 
to the average one shown in Fig. \ref{fig1}(a). 
The onset of the growth spurt of body 2 is later than that of body 1, 
so that the final stature of body 2 is higher than that of body 1. 
Similar result can also be found in the case of the growth of children. 

\begin{figure}[h]
\begin{center}
\includegraphics[width=0.4\textwidth]{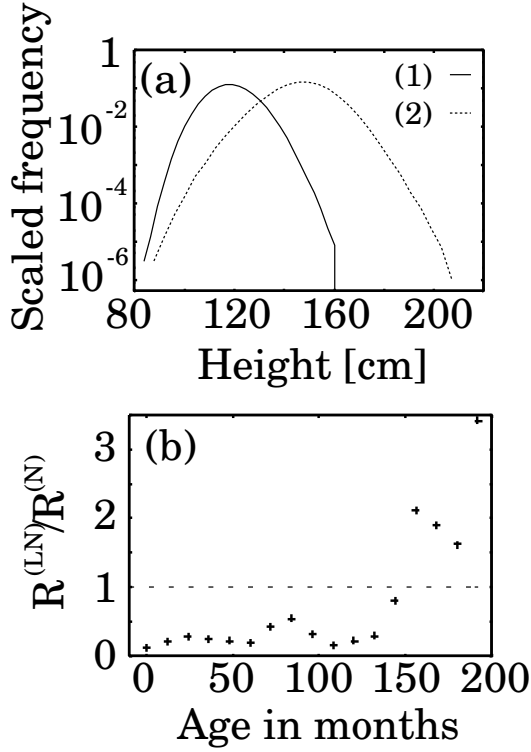}
\end{center}
\caption{
(a) Semi-log plot of height distributions of (1) 96-month-old 
and (2) 156-month-old bodies. 
(b) Relationship between $R^{(LN)}/R^{(N)}$ and age in months.
}
\label{fig5}
\end{figure}

Next, we show the height distribution of 96-month-old and 
156-month-old bodies in Fig. \ref{fig5}(a), where we plot 
the number of bodies scaled by the total number of 
bodies on the vertical axis by the logarithmic scale. 
The height distribution of 
96-month-old bodies (solid curve) looks positively skewed,  
while that of the 156-month-old bodies (dotted curve) shows 
a rather symmetric shape.

Here, let us investigate which statistical distribution 
fits the height distribution better at each age. 
Our procedure of investigation is as follows. 
First,  we fit the height distribution at each age by the 
normal and lognormal distributions using GNUFIT implemented 
in GNUPLOT. 
Next, we calculate the root mean square of residuals, 
\begin{equation}\label{R}
R = \sqrt{\frac{1}{m} \sum_{j=1}^{m} \left( O_{j} - E_{j}\right)^{2}},
\end{equation}
where $O_{j}$ and $E_{j}$ are the frequency and estimated value of 
the fitted distribution of the $j$-th bin, respectively. 
$m$ is the total number of bins.  
Figure \ref{fig5}(b) shows the relationship between the age and 
the ratio of $R^{(LN)}$ to $R^{(N)}$, where $R^{(LN)}$ and $R^{(N)}$ 
are the $R$ values calculated using Eq. (\ref{R}) with the lognormal 
and normal distributions, respectively. 
Figure \ref{fig5}(b) shows that the lognormal 
distribution fits the height distribution well before $156$ months, 
while the normal distribution fits it thereafter. Thus, our growth model 
has succeeded in predicting the transition of the height distribution of 
children found in our previous work. 

\begin{figure}[h]
\begin{center}
\includegraphics[width=0.4\textwidth]{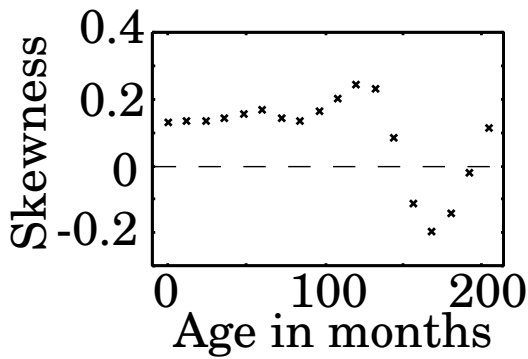}
\end{center}
\caption{
Skewness of height distribution at each age. 
}
\label{fig6}
\end{figure}

In addition, we investigate the skewness of the height distribution defined by 
\begin{equation}\label{skw}
\frac{ <( H(t) - <H(t)> )^{3} >}{[ < ( H(t) - < H(t) > )^{2} >]^{3/2}},
\end{equation}
where the angle brackets denote the ensemble average for all the heights at age $t$. 
Figure \ref{fig6} shows that the skewness changes its sign from positive to negative 
around the onset of a growth spurt followed by a change to a positive value 
at $204$ months old. 
A result similar to this result can also be observed 
in the data analysis of Japanese children\cite{iwata}. 
Except for the data point at $192$ months old in Fig. \ref{fig5}(b), 
the region of the negative skewness 
corresponds to that of $R^{(LN)}/R^{(N)}$ larger than unity, 
which indicates that the change of the sign of the skewness 
may have a close relationship with the transition of the height distribution.

Note that the negative skewness implies that the height distribution 
does not obey the normal distribution rigorously. 
In our previous work, we concluded that the height distribution 
obeys the normal distribution during puberty\cite{kuninaka2009}. 
However, this conclusion was based only on the comparison of the normal and  lognormal 
distributions, so that the normal distribution was regarded 
as the better model for the height distribution during puberty than the lognormal distribution. 
Thus, we will henceforth investigate the reason why the sign of the skewness changes 
during puberty. 
\begin{figure}[h]
\begin{center}
\includegraphics[width=0.4\textwidth]{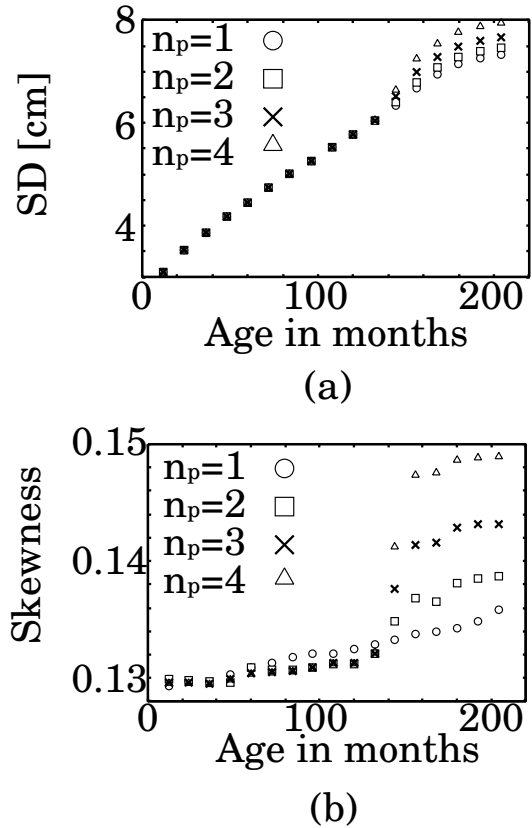}
\end{center}
\caption{
Time evolutions of (a) standard deviation and (b) skewness  
when the onset of growth spurt has no variability. 
}
\label{fig8}
\end{figure}



In our simulation, we have introduced the fluctuation of the growth rate, 
$\delta \alpha^{(i)} (\tilde{t})$, 
by normal random numbers with a constant standard deviation of $\sigma = 10^{-3}$, 
although the value can vary with age in general. 
However, Marubini showed that the individual growth velocity 
during growth spurt (about 24 months) 
has larger fluctuation than those in other periods\cite{marubini}. Thus, 
to investigate the effect of the fluctuation in the growth rate, let us investigate 
the case in which the standard deviation of $\delta \alpha^{(i)} (t)$ depends on time,  
with the onset of growth spurt fixed at $132$ months old. In this simulation, 
we introduce the time-dependent standard deviation $\sigma (t)$ 
of $\delta \alpha^{(i)}(t)$ as 
\begin{equation}
\sigma(t) = 
\begin{cases}
  n_{p} \times 10^{-3} & (132 \le t \le 156)\\
  10^{-3} &  (t < 132,  \hspace{2mm}t > 156),
\end{cases}
\end{equation}
where $n_{p}$ ($\ge 1$) is an integer. 

Figure \ref{fig8}(a) shows the standard deviation against age  
in the cases of $n_{p}=1, 2, 3$, and $4$. Here, we find that the standard 
deviation shows abrupt increases around $144$ months old in the cases of 
$n_{p} = 2, 3, 4$, while it shows an almost linear increase in the case of $n_{p}=1$. 
On the other hand, Fig. \ref{fig8}(b) shows the skewness of the height distribution 
against age, 
where we find that the skewness has larger positive values 
with an increase of $n_{p}$ after $144$ months old. This result indicates 
that the transition of the height distribution does not occur irrespective of $n_{p}$. 
In fact, the ratio $R^{(LN)}/R^{(N)}$ is almost constant at 
about $0.18$ across all ages in the case of $n_{p}=1$, which means 
that the height distribution remains lognormal.  
These results imply that the variability of the onset of the growth spurt is more 
important than the fluctuation in the growth rate for the transition of the height distribution.

\section{Discussion}

In the last section, we have shown that the variability of the onset of the growth 
spurt plays an important role in the transition of the height distribution. 
Here, let us develop a phenomenological argument for explaining the mechanism of the transition. 

As shown in Fig. \ref{fig6}, the skewness of the height distribution changes its sign 
as + $\rightarrow$ - $\rightarrow$ + across the ages. 
Iwata {\it et al.} demonstrated the change of the skewness with their model mimicking 
a growth process of a child by a hyperbolic tangent function of age\cite{iwata}. 
Although their assumption partially includes randomly generated parameters 
that are not based on real data, 
they have succeeded in reproducing the change of the sign of the skewness qualitatively.
Thus, we develop a phenomenological model including parameters estimated from real data. 

Following the model by Iwata {\it et al.}, we mimic the growth of the $i$-th body around its growth 
spurt by the hyperbolic tangent function as
\begin{equation}\label{phe}
H^{(i)}(t) = A^{(i)} + B \tanh \left( \frac{t-t_{0}^{(i)}}{\tau} \right) \hspace{10mm} (96 \le t \le 200).
\end{equation}
Here, $B=19.9$ (cm) and $\tau=35.2$ (months) are constants determined 
by fitting Eq. (\ref{phe}) to the average growth of female children in 2006. 
The parameter $A^{(i)}$ is randomly chosen from the lognormal distribution, Eq. (\ref{ini}),  
with $\mu = 4.846$ and $\sigma = 0.0426$,  which are obtained from the height distribution of 
96-month-old girls in 2006. In addition, the parameter $t_{0}^{(i)}$ is randomly chosen 
from the normal distribution with a mean of $120.5$ (months) and a standard deviation of $\sigma_{v}$. 
From the heights of $10^{6}$ bodies, we calculate the skewness of the height distribution 
at age $t$.

\begin{figure}[h]
\begin{center}
\includegraphics[width=0.4\textwidth]{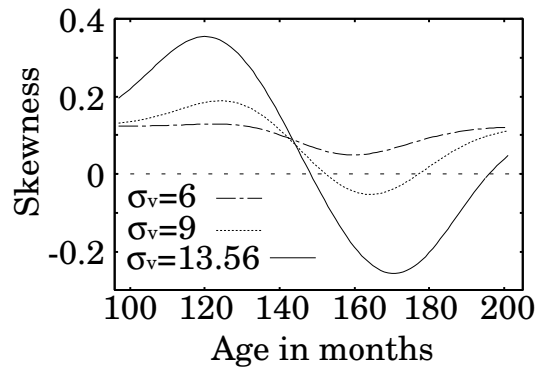}
\end{center}
\caption{
Relationships between age and skewness with different $\sigma_{v}$ values. 
}
\label{fig10}
\end{figure}

Figure \ref{fig10} shows the relationship between the age $t$ and the skewness of the height 
distribution. 
We show three results with $\sigma_{v} = 6$ (chain curve), $\sigma_{v} = 9$ (dotted curve), 
and $\sigma_{v} = 13.56$ (solid curve). 
$\sigma_{v} = 13.56$ is estimated from real data. When $\sigma_{v} < 7.7$, 
the skewness maintains a positive value across all ages, which means that the height distribution 
remains lognormal. Meanwhile, the skewness changes its sign 
as + $\rightarrow$ - $\rightarrow$ + 
when $\sigma_{v} \ge 7.7$, which means that the transition of the height distribution occurs.

\begin{figure}[h]
\begin{center}
\includegraphics[width=0.4\textwidth]{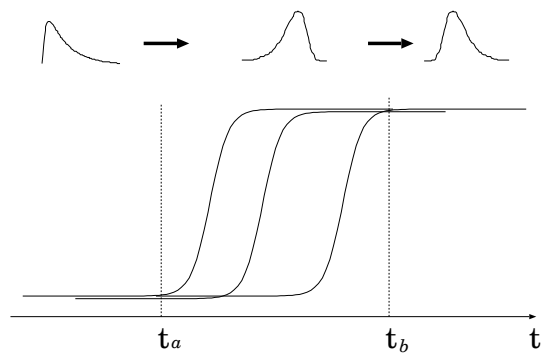}
\end{center}
\caption{
Schematic figure of hyperbolic tangential growths and time evolution 
of height distributions. 
}
\label{schem}
\end{figure}

The mechanism of the change of the skewness can be understood intuitively as follows. 
Figure \ref{schem} shows a schematic figure of the hyperbolic tangential growths 
of three bodies. The change of the height distribution is schematically shown 
in the upper part of Fig. \ref{schem}. Let  $t_a$ be 
the age when any of the bodies starts its growth spurt.  
Before $t_{a}$, the distribution of the heights is lognormal because 
each body grows according to a multiplicative process. 
After $t_{a}$, as the number of tall bodies increases, the peak of the height distribution 
shifts to the right, which results in a negative skewness. 
As the age approaches $t_{b}$ when all the bodies finish their growth,  
bodies with relatively slow growth experience their growth spurt. 
Their final height often become relatively tall 
as stated in the last section. 
Thus, the height distribution will have a longer tail extending to the 
right direction, which results in a positive skewness.

\section{Conclusions}

In this study, we have developed a growth model of children 
that explains the transition of their height distribution 
during puberty, observed in our previous work. Our model is 
based on a multiplicative process, in which the growth rate of 
a body is introduced by adding two kinds of fluctuation 
into the average growth rate of female Japanese  children. 
We have investigated which distribution fits the height distribution 
at each age by calculating the root mean square of residuals 
from the lognomal and normal fitting functions. 
The height distributions fit the normal distribution better 
than the lognormal distribution after 156 months of age 
because the skewness of the height distribution changes its sign during 
puberty. 
For the change of the skewness of the height distribution, 
the variability of the onset of growth spurt is particularly important, as 
demonstrated in our simulation. 
The change of the skewness can be explained by our phenomenological argument, 
which supports the importance of the variability of the onset of growth spurt 
for the transition of the height distribution.



\begin{acknowledgments}
We would like to thank Y. Yamazaki, T. Iwata, M. Katori, 
K. Yamamoto, J. Wakita, and N. Kobayashi for very fruitful discussions. 
HK would like to thank R. Fujie, S. Mori, M. Tanaka, S. Matsushita, 
and R. Murakami for their valuable comments. 
This work is supported (in part) by a Grant for Basic Science Research Projects 
from The Sumitomo Foundation. 
\end{acknowledgments}


\end{document}